\documentstyle[prd,aps,eqsecnum,preprint,tighten]{revtex}
\begin{document}
\title{Selection Rules in Minisuperspace Quantum Cosmology}
\author{S. Capozziello\thanks{E-mail: capozziello@physics.unisa.it}~~ and
G. Lambiase\thanks{E-mail: lambiase@physics.unisa.it}}
\address{Dipartimento di Scienze Fisiche "E.R. Caianiello" \\
 Universit\'a di Salerno, 84081 Baronissi (Sa), Italy. \\
 Istituto Nazionale di Fisica Nucleare, Sez. di Napoli, Italy. }
\date{\today}
\maketitle
\begin{abstract}
The existence of a Noether symmetry for a given minisuperspace
cosmological model is a sort of selection rule to recover
classical behaviours in cosmic evolution since oscillatory
regimes for the wave function of the universe come out.
The so called Hartle criterion to select correlated regions in
the configuration space of dynamical variables can be directly
connected to the presence of a Noether symmetry and we show that
such a statement works for generic extended theories of gravity
in the framework of minisuperspace approximation.
Examples and exact cosmological solutions are given for
nonminimally coupled and higher--order theories.
\end{abstract}
\thispagestyle{empty} 

\section{Introduction}

Several points of view can be adopted in order to define quantum
cosmology. It can be considered as the first step toward the
construction of a full theory of quantum gravity. Besides, it
concerns finding initial conditions from which our {\it classical}
universe is started. However, with respect to other theories of
physics as electromagnetism, general relativity or ordinary
quantum mechanics, boundary conditions for the evolution of the
system {\it ``universe''} cannot be set from {\it outside}. There we
need a fundamental dynamical law (e.g. Maxwell's or Einstein's
equations or  Schr\"odinger's equation) and then we {\it impose},
from the outside, the initial conditions. In cosmology, by
definition, there is no {\it rest of the universe} so that
boundary conditions must be a {\it fundamental law of physics}. In
this sense, a part the fact that quantum cosmology is a workable
scheme to achieve quantum gravity, it can be considered as an
autonomous branch of physics due to the issue of finding initial
conditions \cite{hartle1}.

However, not only the conceptual difficulties, but also
mathematical ones make quantum cosmology hard to handle. For
example, the superspace of geometrodynamics \cite{wheeler} has
infinite degrees of freedom so that it is practically impossible
to integrate the full Wheeler--DeWitt (WDW) equation. Furthermore,
a Hilbert space of states describing the universe is not
available \cite{deWitt}. Finally, it is not clear how to interpret
the solutions of WDW equation in the framework of probability
theory. Several interpretative schemes have been proposed but the
concepts of {\it probability} and {\it unitarity} are in any case
approximate. Their validity is limited by the accuracy of the
semiclassical approximation and strictly depends on the suitable
definition of probability current \cite{deWitt},\cite{vilenkin}.

Despite these still unsolved shortcomings, several positive
results have been obtained and quantum cosmology has become a sort
of {\it paradigm} in theoretical physics researches. For example
the infinite--dimensional superspace can be restricted to
opportune finite--dimensional configuration spaces called {\it
minisuperspaces}. In this case, the above mathematical difficulties
can be avoided
and the WDW equation can be integrated.
 The so called {\it no boundary condition} by Harte and Hawking
\cite{HH} and the {\it tunneling from nothing} by Vilenkin
\cite{vilenkin} give reasonable laws for initial conditions from
which our {\it classical} universe could be started.

The {\it Hartle criterion} \cite{hartle2} is an interpretative scheme for
the solutions of the WDW equation. Hartle proposed to look for peaks
of the wave function of the universe: If it is strongly peaked,
we have correlations among the geometrical
and matter degrees of freedom; if it is not peaked, correlations
are lost. In the first case, the emergence of classical
trajectories (i.e. universes) is expected. The analogy to the
non--relativistic quantum mechanics is immediate. If we have a
potential barrier and a wave function, solution of the
Schr\"odinger equation, we have an oscillatory regime on and
outside the barrier; we have a decreasing exponential behaviour
under the barrier.
The system behaves classically in the oscillatory regime while it does not in
the exponential case.
The situation is analogous in quantum
cosmology: Now the potential barrier has to be replaced by the
superpotential $U(h^{ij}, \varphi)$, where $h^{ij}$ are the components of the
three--metric of geometrodynamics and $\varphi$ is a generic
scalar field describing the matter content. More precisely, the
wave function of the universe can be written as
 \begin{equation}\label{1}
 \Psi[h_{ij}(x), \phi(x)]\sim e^{im_P^2S}\,{,}
 \end{equation}
where $m_p$ is the Planck mass and $S$ is an action. A state with  classical
correlations must be a superposition of states of the form (\ref{1}).
This type of state can be expressed as a coherent superposition of eigenstates
of operators that commute with the constraints and correspond to constants
of the motion. A superposition of this kind can be approximated by a WKB
state where
 \begin{equation}\label{2}
 S\equiv S_0+m_P^{-2}S_1+O(m_P^{-4})\,{,}
 \end{equation}
 is the expansion of the action.
We have to note that there is no normalization factor due to the lack
of a probability interpretative scheme.
However, in this approximation, it is possible to define a localized
prefactor as shown in \cite{lifschytz} and we can define a
{\it quasiclassical} state to describe an approximate classical
behaviour and a {\it semiclassical} one to describe a product of a part
that is quasiclassical and a part that is not. This quasiclassical state
can be a coherent one if it is a superposition of states in the sense discussed
in \cite{lifschytz}. Transition amplitudes of istantaneous eigenstates are
discussed in \cite{massar}.

Considering the action (\ref{2}) and inserting it into the WDW
equation and equating similar power of $m_p$, one obtains the
Hamilton--Jacobi equation for $S_0$. Similarly, one gets equations
for $S_1, S_2,\ldots$, which can be solved considering results of
previous orders. We need only $S_0$ to recover the
semi--classical limit of quantum cosmology \cite{halliwell}. If
$S_0$ is a real number, we get oscillating WKB modes and
$\Psi$ is peaked on a
phase--space region defined by
 \begin{equation}\label{3}
 \pi_{ij}=m_P^2\,\frac{\delta S_0}{\delta h^{ij}}\,{,}\quad
 \pi_{\varphi}=m_P^2\,\frac{\delta S_0}{\delta\varphi}\,{,}
 \end{equation}
where $\pi_{ij}$ and $\pi_{\varphi}$ are classical momenta
conjugates to $h^{ij}$ and $\varphi$. The semi--classical region
of superspace, where $\Psi$ has an oscillating structure, is the
Lorentz one otherwise it is Euclidean. In the latter case, we have
$S=iI$ and
 \begin{equation}\label{4}
 \Psi\sim e^{-m_P^2 I}\,{,}
 \end{equation}
where $I$ is the action for the Euclidean solutions of classical field
equations ({\it istantons}).
Given an action $S_0$, Eqs.(\ref{3}) imply $n$ free parameters
(one for each dimension of the configuration space ${\cal Q}\equiv\{h^{ij},
\varphi\}$) and then $n$ first integrals of motion. However the
general solution of the field equations involves $2n-1$ parameters
(one for each Hamilton equation of motion except the energy
constraint). Consequently, the wave function oscillates on a subset
of the general solution. In this sense, the boundary conditions on
the wave function (e.g. Harte--Hawking, Vilenkin, others) imply
initial conditions for the classical solutions.

To be more precise, if the wave function $\Psi$
is sufficiently peaked about some region in configuration space,
we predict that we will observe the correlations between observables
which characterize this region. If $\Psi$ is small in some region,
we predict that observations of the correlations which characterize
this region are precluded. Where $\Psi$ is neither small nor sufficiently
peaked, we do not predict anything.

For example, given the measured value of the Hubble constant and mass density,
we would like a "good" wave function for the universe to be peaked around
a distribution of galaxies consistent with that which is observed.
It is crucial to recognize that the wave function does not predict
a specific value for $H_{0}$, or specific locations for the galaxies,
but rather a "correlation" between these observables.

Halliwell \cite{halliwell} has shown that an oscillatory wave function
of the form (\ref{1}) predicts a correlation between the canonical
coordinate $q$ and the momentum $\pi_{q}$ of the above form
$\pi_{q}=m_{P}^2 \partial S/\partial q$.
In other words, taking into account minisuperspace models, if oscillatory
regimes of the WDW wave function exist, we are able to recover correlations
among variables and then classical behaviours emerge.
Again, in the classically allowed region,
the semiclassical approximation to the
WDW wave function yields just such  oscillatory solutions.

A simple minisuperspace example can be constructed.
The ansatz
for the wave function is
\begin{equation}
\label{1a}
\Psi(a)=e^{i m_{P}^2S(a)}\,,
\end{equation}
where the canonical variables $q$ coincides to the scale factor of the universe
$a$ and the phase is a slowly varying function of the scale factor. Since we
wish to investigate the classical limit $m_P^2\to\infty$ (which corresponds
to $\hbar\to 0$ of ordinary quantum mechanics), we use the expansion (\ref{2}).
Inserting such an ansatz into the WDW equation
\begin{equation}
\label{2a}
\left[ \frac{\partial^2}{\partial a^2}-\left(\frac{3\pi}{2G}\right)^2
a^2\left(1-\frac{a^2}{a_0^2}\right)\right]\Psi (a)=0\,,
\end{equation}
deduced by the action
\begin{equation}
\label{3a}
{\cal A}=\frac{3\pi}{4G}\int dt \left[-\dot{a}^2a+a
\left(1-\frac{a^2}{a_0^2}\right)\right]\,,
\end{equation}
of a Friedman--Robertson--Walker (FRW) closed universe ($a_{0}$ is a constant)
with the
canonical momentum $\pi_{a}$ given by
\begin{equation}
\label{4a}
\pi_{a}=-\frac{3\pi}{2G}a\dot{a}\,,
\end{equation}
we get a set of differential equations, one for any order of
$m_{P}^{-2}$, which as we said above, can be solved sequentially.
The semiclassical approximation to the wave function obtains by
working only to first order. We get
\begin{equation}
\label{5a}
S_{0}=\int^a da'\sqrt{\left(\frac{3\pi}{2G}\right)^2 a^{2}_{0}
\left(\frac{a^4}{a_0^4}-\frac{a^2}{a_0^2}\right)}\,,
\end{equation}
and
\begin{equation}
\label{6a}
S_{1}=\frac{1}{2}\ln\left(\frac{\partial S_0}{\partial a}\right)\,.
\end{equation}
Thus the oscillatory semiclassical wave function
$\Psi\propto \exp(iS_{0})$ is peaked about a region of minisuperspace
(every point of which represents a closed FRW model)
in which the correlation between the coordinate and momentum (scale factor and
expansion rate), $\pi_a=\partial S_0/\partial a$, holds good.

Using Eq. (\ref{4a}) for $\pi_a$, the correlation reduces to
\begin{equation}
\label{7a}
\dot{a}=\sqrt{\frac{a^2}{a_0^2}-1}\,,
\end{equation}
which is nothing else but the $(0,0)$ Einstein equation for a FRW
spacetime. If
\begin{equation}\label{8a}
a_0=\sqrt{\frac{3}{\Lambda}}\,, \quad \Lambda=8\pi G\rho_{vac}\,,
\end{equation}
where $\rho_{vac}$ is a constant density, we get the solution
\begin{equation}
\label{9a}
a(t)=a_0\cosh (a_0^{-1}t)\,,
\end{equation}
which is an inflationary behaviour for a closed FRW model. Thus, by this
simple example, in the region of minisuperspace where the wave function
 oscillates, a classical FRW spacetime, obeying the (classical)
Einstein equation emerges.

The issue is now if there exists some method capable of selecting
such constants of motion which, being first integrals of motion,
allow to find correlations between classical variables and conjugate momenta
in  minisuperspace models. In other words,
can the emergence of classical trajectories be implemented by
some general approach without arbitrarily choosing regions of the
phase--space where momenta (\ref{3}) are constant?
Achieving this result means to obtain oscillatory subsets of WDW wave
function where one gets correlations. Consequently classical regime
are recovered and the Hartle criterion holds (at least in the framework
of the minisuperspace approximation). For the full theory, i.e. without
considering simple minisuperspace models,
the issue is more delicate since we have to ask
for superpositions of the form (\ref{1}) which yield peaked wave packets
so that the Hartle criterion holds \cite{lifschytz}. In this case,
also the issue of universe ``creation'', as a particle creation problem,
has to be faced considering the way in which the quantum--classical
transition is achieved \cite{massar}.

In this paper, we want to restrict to a more specific (and simple)
 question. We want
to show, for general extended gravity minisuperspace models,
that the existence of
a Noether symmetry implies, at least, a subset of the general
solution of the WDW equation where the oscillating behaviour is
recovered. Viceversa,
the presence of a Noether symmetry gives rise  to the emergence of
classical trajectories. This analysis is performed in the context
of the minisuperspace approximation and,
for {\it classical trajectories}, we mean
solutions of the ordinary Einstein equations.
The existence of a Noether symmetry for a dynamical model
is a general criterion to search for constants (first integrals) of motion
so that, given a minisuperspace model exhibiting such symmetries we
obtain certainly  correlations and then classical behaviours.
This statement, in our knowledge, has never been done also if constants
of motion have been systematically used in quantum cosmology since at least
fifteen years.

The layout of the paper is the following. Sect. 2 is devoted to
the Noether Symmetry Approach and to its connection to quantum
cosmology. In Sect. 3, we apply the method to minisuperspace
models of nonminimally coupled theories of gravity, while the same
is done in Sects. 4 and 5 for higher--order theories. Discussion
and conclusions are drawn in Sect. 6.

\section{The Noether Symmetry Approach and Quantum Cosmology}

Minisuperspaces are restrictions of the superspace of
geometrodynamics. They are finite--dimensional configuration spaces
on which point-like Lagrangians can be defined.
Cosmological models of physical interest can be defined on such
minisuperspaces (e.g. Bianchi models).

Before taking into account specific models, let us remind some
properties of the Lie derivative and the derivation of the
Noether theorem \cite{arnold}.
 Let $L_X$ be the Lie derivative
 \begin{equation}\label{5}
 (L_X\omega)\xi=\frac{d}{dt}\omega(g_*^t\xi)\,{,}
 \end{equation}
 where $\omega$ is a differential form of ${\cal R}^n$ defined on the
vector field $\xi$, $g_*^t$ is the differential of the phase flux
$\{g_t\}$ given by the vector field $X$ on a differential manifold
${\cal M}$. Let $\rho_t=\rho_{g-t}$ be the action of a one--parameter
group able to act on functions, vectors and forms on the vector
spaces $C^{\infty}({\cal M})$, $D({\cal M})$, and $\Lambda({\cal M})$
constructed starting from ${\cal M}$.
 If $g_t$ takes the point $m\in M$ in $g_t(m)$, then $\rho_t$
takes back on $m$ the vectors and the forms defined on $g_t(m)$;
$\rho_t$ is a {\it pull back} \cite{marmo}. Then the property
 \begin{equation}\label{6}
 \rho_{t+s}=\rho_t\rho_s
 \end{equation}
 holds since
 \begin{equation}\label{7}
 g_{t+s}=g_t\circ g_s\,{.}
 \end{equation}
 On the functions $f, g\in {\cal C}^{\infty}({\cal M})$ we have
 \begin{equation}\label{8}
 \rho_t(fg)=(\rho_tf)(\rho_tg)\,{;}
 \end{equation}
 on the vectors $X, Y\in {\cal D}({\cal M})$,
 \begin{equation}\label{9}
 \rho_t[X, Y]=[\rho_tX, \rho_tY]\,{;}
 \end{equation}
 on the forms $\omega, \mu\in \Lambda({\cal M})$
 \begin{equation}\label{10}
 \rho_t(\omega\wedge \mu)=(\rho_t\omega)\wedge (\rho_t\mu)\,{.}
 \end{equation}
 $L_X$ is the infinitesimal generator of the one--parameter
group $\rho_t$, and, being a derivative on the algebras
${\cal C}^{\infty}({\cal M})$, ${\cal D}({\cal M})$, and $\Lambda({\cal M})$,
the following properties have to hold
 \begin{eqnarray}
 L_X(fg)&=&(L_Xf)g+f(L_Xg)\,{,} \label{11} \\
 L_X[Y, Z]&=&[L_XY, Z]+[Y, L_XZ]\,{,}\label{12}\\
 L_X(\omega\wedge\mu)&=&(L_X\omega)\wedge\mu+\omega\wedge(L_X\mu)\,{,}\label{13}
 \end{eqnarray}
 which are nothing else but the Leibniz rules for  functions,
 vectors and  differential forms, respectively. Furthermore,
 \begin{eqnarray}
 L_Xf&=&Xf\,{,} \label{14} \\
 L_XY&=&adX(Y)=[X, Y]\,{,}\label{15} \\
 L_Xd\omega&=&dL_X\omega\,{,}\label{16}
 \end{eqnarray}
 where $ad$ is the self--adjoint operator and $d$ is the external
derivative by which a $p$--form becomes a $(p+1)$--form.

The discussion can be specified by considering a Lagrangian ${\cal
L}$ which is a function defined on the tangent space of
configurations ${\cal TQ}\equiv\{q_i, \dot{q}_i\}$. In this case, the vector
field $X$ is
 \begin{equation}\label{17}
 X=\alpha^i(q)\frac{\partial}{\partial q^i}+
 \dot{\alpha}^i(q)\frac{\partial}{\partial\dot{q}^i}\,{,}
 \end{equation}
 where dot means derivative with respect to $t$, and
 \begin{equation}\label{18}
 L_X{\cal L}=X{\cal L}=\alpha^i(q)\frac{\partial {\cal L}}{\partial q^i}+
 \dot{\alpha}^i(q)\frac{\partial {\cal L}}{\partial\dot{q}^i}\,{.}
 \end{equation}
 The condition
 \begin{equation}\label{19}
 L_X{\cal L}=0
 \end{equation}
 implies that the phase flux is conserved along $X$: This means that a
constant of motion exists for ${\cal L}$ and the Noether theorem
holds. In fact, taking into account the Euler--Lagrange equations
 \begin{equation}\label{20}
 \frac{d}{dt}\frac{\partial {\cal L}}{\partial\dot{q}^i}-
 \frac{\partial {\cal L}}{\partial q^i}=0\,{,}
 \end{equation}
 it is easy to show that
 \begin{equation}\label{21}
 \frac{d}{dt}\left(\alpha^i\frac{\partial {\cal
 L}}{\partial\dot{q}^i}\right)=L_X{\cal L}\,{.}
 \end{equation}
 If (\ref{19}) holds,
 \begin{equation}\label{22}
 \Sigma_0=\alpha^i\frac{\partial {\cal L}}{\partial\dot{q}^i}
 \end{equation}
 is a constant of motion. Alternatively, using the Cartan
one--form
 \begin{equation}\label{23}
 \theta_{{\cal L}}\equiv\frac{\partial {\cal
 L}}{\partial\dot{q}^i}\,dq^i
 \end{equation}
 and defining the inner derivative
 \begin{equation}\label{24}
 i_X\theta_{{\cal L}}=<\theta_{{\cal L}}, X>\,{,}
 \end{equation}
 we get, as above,
 \begin{equation}\label{25}
 i_X\theta_{{\cal L}}=\Sigma_0
 \end{equation}
 if condition (\ref{19}) holds.
 This representation is useful to identify cyclic variables. Using
a point transformation on vector field (\ref{17}), it is possible
to get\footnote{We shall indicate the quantities as
Lagrangians and  vector fields with a tilde if the
non--degenerate transformation
 $$
 Q^i=Q^i(q)\,{,}\quad \dot{Q}^i(q)=\frac{\partial Q^i}{\partial
 q^j}\,\dot{q}^j
 $$
 is performed. However the Jacobian determinant
 ${\cal J}=\parallel\partial Q^i
 /\partial q^j\parallel$ has to be non--zero.}
 \begin{equation}\label{26}
 \tilde{X}=(i_XdQ^k)\,\frac{\partial}{\partial Q^k}+
 \left[\frac{d}{dt}(i_XdQ^k)\right]\frac{\partial}{\partial\dot{Q}^k}\,{.}
 \end{equation}
 If $X$ is a symmetry also $\tilde{X}$ has this property, then it
is always possible to choose a coordinate transformation so that
 \begin{equation}\label{27}
 i_XdQ^1=1\,{,} \quad i_XdQ^i=0\,{,}\quad i\neq 1\,{,}
 \end{equation}
 and then
 \begin{equation}\label{28}
 \tilde{X}=\frac{\partial}{\partial Q^1}\,{,}\quad
 \frac{\partial\tilde{{\cal L}}}{\partial Q^1}=0\,{.}
 \end{equation}
 It is evident that $Q^1$ is the cyclic coordinate and the
dynamics can be reduced \cite{arnold}. However, the change of
coordinates is not unique and a clever choice is always important.
 Furthermore, it is possible that more symmetries are found.
In this case more cyclic variables exists. For example, if $X_1,
X_2$ are the Noether vector fields and they commute, $[X_1,
X_2]=0$, we obtain two cyclic coordinates by solving the system
 \begin{equation}\label{29}
 i_{X_1}dQ^1=1\,{,}\quad i_{X_2}dQ^2=1\,{,}
 \end{equation}
 $$
 i_{X_1}dQ^i=0\,{,}\quad i\neq 1\,{;} \quad  i_{X_2}dQ^i=0\,{,}\quad
 i\neq 2\,{.}
 $$
 If they do not commute, this procedure does not work since
commutation relations are preserved by diffeomorphisms. In this
case
 \begin{equation}\label{30}
 X_3=[X_1, X_2]
 \end{equation}
 is again a symmetry since
 \begin{equation}\label{31}
 L_{X_3}{\cal L}=L_{X_1}L_{X_2}{\cal L}-L_{X_2}L_{X_1}{\cal
 L}=0\,{.}
 \end{equation}
 If $X_3$ is independent of $X_1, X_2$ we can go on until the
vector fields close the Lie algebra \cite{bianchi}.

A reduction procedure by cyclic coordinates can be implemented in
three steps: i) we choose a symmetry and obtain new coordinates as
above. After this first reduction, we get a new Lagrangian
$\tilde{{\cal L}}$ with a cyclic coordinate; ii) we search for new
symmetries in this new space and apply the reduction technique
until it is possible; iii) the process stops if we select a pure
kinetic Lagrangian where all coordinates are cyclic. This case is
not very common and often it is not physically relevant. Going
back to the point of view interesting in quantum cosmology, any
symmetry selects a constant conjugate momentum since, by the
Euler--Lagrange equations
 \begin{equation}\label{32}
 \frac{\partial\tilde{{\cal L}}}{\partial Q^i}=0\Longleftrightarrow
 \frac{\partial\tilde{{\cal L}}}{\partial \dot{Q}^i}=\Sigma_i\,{.}
 \end{equation}
 Viceversa, the existence of a constant conjugate momentum means
that a cyclic variable has to exist. In other words, a Noether
symmetry exists.

Further remarks on the form of the Lagrangian ${\cal L}$ are
necessary at this point. We shall take into account
time--independent, non--degenerate Lagrangians ${\cal L}={\cal
L}(q^i, \dot{q}^j)$, i.e.
 \begin{equation}\label{33}
 \frac{\partial {\cal L}}{\partial t}=0\,{,}\quad
 \mbox{det} H_{ij}\equiv \mbox{det}\vert\vert\frac{\partial^2{\cal L}}
 {\partial\dot{q}^i\partial\dot{q}^j}\vert\vert \neq 0\,{,}
 \end{equation}
 where $H_{ij}$ is the Hessian.
 As in usual analytic mechanics, ${\cal L}$ can be set in the form
 \begin{equation}\label{34}
 {\cal L}=T(q^i, \dot{q}^i)-V(q^i)\,{,}
 \end{equation}
 where $T$ is a positive--defined quadratic form in the
$\dot{q}^j$ and $V(q^i)$ is a potential term. The energy function
associated with ${\cal L}$ is
 \begin{equation}\label{35}
 E_{{\cal L}}\equiv \frac{\partial {\cal L} }{\partial
 \dot{q}^i}\,\dot{q}^i-{\cal L}(q^j, \dot{q}^j)
 \end{equation}
 and by the Legendre transformations
 \begin{equation}\label{36}
 {\cal H}=\pi_j\dot{q}^j-{\cal L}(q^j, \dot{q}^j)\,{,}\quad
 \pi_j=\frac{\partial {\cal L}}{\partial \dot{q}^j}\,{,}
 \end{equation}
 we get the Hamiltonian function and the conjugate momenta.

Considering again the symmetry, the condition (\ref{19}) and the
 vector field $X$ in Eq.(\ref{17}) give a homogeneous polynomial
of second degree in the velocities plus an inhomogeneous term in
the $q^j$. Due to (\ref{19}), such a polynomial has to be
identically zero and then each coefficient must be independently
zero. If $n$ is the dimension of the configuration space (i.e. the
dimension of the minisuperspace), we get $\{ 1+n(n+1)/2\}$ partial
differential equations whose solutions assign the symmetry, as we
shall see below. Such a symmetry is over--determined and, if a
solution exists, it is expressed in terms of integration constants
instead of boundary conditions.

In the Hamiltonian formalism, we have
 \begin{equation}\label{43}
 [\Sigma_j, {\cal H}]=0\,{,}\quad 1\leq j \leq m\,{,}
 \end{equation}
 as it must be for conserved momenta in quantum mechanics and the
Hamiltonian has to satisfy the relations
 \begin{equation}\label{44}
 L_{\Gamma}{\cal H}=0\,{,}
 \end{equation}
in order to obtain a Noether symmetry. The vector $\Gamma$ is defined by
\cite{marmo}
 \begin{equation}\label{45}
 \Gamma =\dot{q}^i\frac{\partial}{\partial q^i}+
 \ddot{q}^i\frac{\partial}{\partial\dot{q}^i}\,{.}
 \end{equation}
Let us now go to the minisuperspace quantum cosmology and to the
semi--classical interpretation of the wave function of the
universe.

By a straightforward canonical quantization procedure, we have
 \begin{eqnarray}
 \pi_j& \longrightarrow & \hat{\pi}_j=-i\partial_j\,{,} \label{37} \\
 {\cal H} & \longrightarrow & \hat{\cal H}(q^j,
 -i\partial_{q^j})\,{.}\label{38}
 \end{eqnarray}
It is well know that the Hamiltonian constraint gives the WDW
equation, so that if $\vert \Psi>$ is a {\it state} of the system
(i.e. the wave function of the universe), dynamics is given by
 \begin{equation}\label{39}
 {\cal H}\vert\Psi>=0\,{.}
 \end{equation}
 If a Noether symmetry exists, the reduction procedure outlined
above can be applied and then, from (\ref{32}) and (\ref{36}), we
get
 \begin{eqnarray}
 \pi_1\equiv\frac{\partial {\cal
 L}}{\partial\dot{Q}^1}=i_{X_1}\theta_{{\cal L}}& = &
 \Sigma_1\,{,}\nonumber \\
 \pi_2\equiv\frac{\partial {\cal
 L}}{\partial\dot{Q}^2}=i_{X_2}\theta_{{\cal L}}& = &
 \Sigma_2\,{,}\label{40} \\
\ldots\quad \ldots & & \ldots \,{,} \nonumber
 \end{eqnarray}
 depending on the number of Noether symmetries. After
quantization, we get
 \begin{eqnarray}
 -i\partial_1\vert\Psi>&=&\Sigma_1\vert\Psi>\,{,} \nonumber \\
 -i\partial_2\vert\Psi>&=&\Sigma_2\vert\Psi>\,{,} \label{41} \\
 \ldots & & \ldots \,{,} \nonumber
 \end{eqnarray}
which are nothing else but translations along the $Q^j$ axis
singled out by corresponding symmetry. Eqs. (\ref{41}) can be
immediately integrated and, being $\Sigma_j$ real constants, we
obtain oscillatory behaviours for $\vert \Psi>$ in the directions
of symmetries, i.e.
 \begin{equation}\label{42}
 \vert\Psi>=\sum_{j=1}^m\, e^{i\Sigma_jQ^j}\vert \chi(Q^l)>\,{,}
 \quad m < l\leq n\,{,}
 \end{equation}
 where $m$ is the number of symmetries, $l$ are the directions
where symmetries do not exist, $n$ is the total dimension of
minisuperspace. It is worthwhile to note that the component
$|\chi>$ of the wave function could also depend on $\Sigma_{j}$
but it is not possible to state ``in general'' if it is oscillating.

Viceversa, dynamics given by (\ref{39}) can be reduced by
(\ref{41}) if and only if it is possible to define constant
conjugate momenta as in (\ref{40}), that is oscillatory behaviours
of a subset of solutions $\vert\Psi>$ exist only if Noether
symmetry exists for dynamics.

The $m$ symmetries give first integrals of motion and then the
possibility to select classical trajectories. In one and
two--dimensional minisuperspaces, the existence of a Noether
symmetry allows the complete solution of the problem and to get
the full semi--classical limit of minisuperspace quantum cosmology.
By these arguments, the Halliwell request that an oscillatory wave function
predict correlations between  coordinates and canonical conjugate momenta
\cite{halliwell} is fully recovered.

In conclusion, we can set out the following

\vspace{1.5 cm}

\noindent {\bf Theorem}: {\it In the semi--classical limit of quantum
cosmology and in the framework of minisuperspace approximation,
the reduction procedure of dynamics, due to the
existence of Noether symmetries, allows to select a subset of the
solution of WDW equation where oscillatory behaviours are found.
As consequence, correlations between coordinates and canonical conjugate
momenta emerge so that
classical cosmological solutions can be recovered.
Viceversa, if a subset of the solution of WDW equation has an
oscillatory behaviour, due to Eq.(\ref{41}), conserved momenta have to
exist and Noether symmetries are present. In other words, Noether
symmetries select classical universes.}

\vspace{1.5 cm}

\noindent In what follows, we shall give realizations of such a statement
for minisuperspace cosmological models derived from extended
gravity theories.

\section{Scalar--Tensor Gravity Cosmologies}

Let us take into account a nonminimally coupled theory of gravity
of the form
 \begin{equation}\label{46}
 {\cal A}=\int d^4x\sqrt{-g}\left[F(\varphi)R+\frac{1}{2}g^{\mu\nu}
 \varphi_{\mu}\varphi_{\nu}-V(\varphi)\right]\,{,}
 \end{equation}
 where $F(\varphi)$ and $V(\varphi)$ are respectively the coupling
and the potential of a scalar field \cite{pla}. We are using, from now on,
physical units $8\pi G=c=\hbar =1$, so that the standard Einstein
coupling is recovered for $F(\varphi)=-1/2$.

Let us restrict, for the sake of simplicity, to a
FRW cosmology. The Lagrangian in
(\ref{46}) becomes
 \begin{equation}\label{47}
 {\cal L}=6a\dot{a}^2F+6a^2\dot{a}\dot{F}-
 6kaF+a^3\left[\frac{\dot{\varphi}}{2}-V\right]\,{,}
 \end{equation}
 in terms of the scale factor $a$.

The configuration space of such a Lagrangian is ${\cal Q}\equiv\{a, \varphi\}$, i.e. a
two--dimensional minisuperspace. A Noether symmetry exists if (\ref{19}) holds. In this
case, it has to be
 \begin{equation}\label{48}
 X=\alpha\, \frac{\partial}{\partial a}+
 \beta\,\frac{\partial}{\partial \varphi}+
 \dot{\alpha}\,\frac{\partial}{\partial\dot{a}}+
 \dot{\beta}\,\frac{\partial}{\partial\dot{\varphi}}\,{,}
 \end{equation}
where $\alpha, \beta$ depend on $a, \varphi$. The system of
partial differential equation given by (\ref{19}) is
 \begin{eqnarray}
&& F(\varphi)\left[\alpha+2a\frac{\partial\alpha}{\partial
 a}\right]+
 a F'(\varphi)\left[\beta+a\frac{\partial\beta}{\partial
 a}\right] =  0\,{,} \label{49}\\
&& 3\alpha+12F'(\varphi)\frac{\partial\alpha}{\partial\varphi}+2a
 \frac{\partial\beta}{\partial\varphi}  =  0\,{,}\label{50} \\
&& a\beta F''(\varphi)+\left[2\alpha+a\frac{\partial \alpha}{\partial a}
 +\frac{\partial\beta}{\partial\varphi}\right]F'(\varphi)+
 2\frac{\partial\alpha}{\partial\varphi}F(\varphi)+\frac{a^2}{6}
 \frac{\partial\beta}{\partial a} =  0\,{,} \label{51} \\
&& [3\alpha V(\varphi)+a\beta V'(\varphi)]a^2+6k[\alpha F(\varphi)+
 a\beta F'(\varphi)]  =  0\,{.}\label{52}
 \end{eqnarray}
The prime indicates the derivative with respect to $\varphi$. The
number of equations is $4$ as it has to be, being $n=2$. Several
solutions exist for this system \cite{pla,cqg,prd}. They determine
also the form of the model since the system (\ref{49})--(\ref{52})
gives $\alpha, \beta$, $F(\varphi)$ and $V(\varphi)$. For example,
if the spatial curvature is $k=0$, a solution is
 \begin{equation}\label{53}
 \alpha=-\frac{2}{3}p(s)\beta_0a^{s+1}\varphi^{m(s)-1}\,{,}\quad
 \beta=\beta_0a^s\varphi^{m(s)}\,{,}
 \end{equation}
 \begin{equation}\label{54}
 F(\varphi)=D(s)\varphi^2\,{,}\quad
 V(\varphi)=\lambda\varphi^{2p(s)}\,{,}
 \end{equation}
 where
 \begin{equation}\label{55}
 D(s)=\frac{(2s+3)^2}{48(s+1)(s+2)}\,{,}\quad
 p(s)=\frac{3(s+1)}{2s+3}\,{,}\quad
 m(s)=\frac{2s^2+6s+3}{2s+3}\,{,}
 \end{equation}
 and $s, \lambda$ are free parameters. The change of variables
(\ref{27}) gives
 \begin{equation}\label{56}
 w=\sigma_0a^3\varphi^{2p(s)}\,{,}\quad z=\frac{3}{\beta_0\chi(s)}
 a^{-s}\varphi^{1-m(s)}\,{,}
 \end{equation}
 where $\sigma_0$ is an integration constant and
 \begin{equation}\label{57}
 \chi(s)=-\frac{6s}{2s+3}\,{.}
 \end{equation}
 Lagrangian (\ref{47}) becomes, for $k=0$,
 \begin{equation}\label{58}
 {\cal L}=\gamma(s)w^{s/3}\dot{z}\dot{w}-\lambda w\,{,}
 \end{equation}
 where $z$ is cyclic and
 \begin{equation}\label{59}
 \gamma(s)=\frac{2s+3}{12\sigma_0^2(s+2)(s+1)}\,{.}
 \end{equation}
 The conjugate momenta are
 \begin{equation}\label{60}
 \pi_z=\frac{\partial {\cal L}}{\partial \dot{z}}=\gamma(s)w^{s/3}
 \dot{w}\,{,}\quad
 \pi_w=\frac{\partial {\cal L}}{\partial \dot{w}}=\gamma(s)w^{s/3}
 \dot{z}\,{,}
 \end{equation}
 and the Hamiltonian is
 \begin{equation}\label{61}
 \tilde{{\cal
 H}}=\frac{\pi_z\pi_w}{\gamma(s)w^{s/3}}+\lambda w\,{.}
 \end{equation}
 The Noether symmetry is given by
 \begin{equation}\label{62}
 \pi_z=\Sigma_0\,{.}
 \end{equation}
 Quantizing Eqs. (\ref{60}), we have
 \begin{equation}\label{63}
 \pi\longrightarrow -i\partial_z \,{,}\quad \pi_w\longrightarrow
 -i\partial_w\,{,}
 \end{equation}
 and then the WDW equation
 \begin{equation}\label{64}
 [(i\partial_z)(i\partial_w)+\tilde{\lambda}w^{1+s/3}]\vert\Psi>=0\,{,}
 \end{equation}
 where $\tilde{\lambda}=\gamma(s)\lambda$.

The quantum version of constraint (\ref{62}) is
 \begin{equation}\label{65}
 -i\partial_z\vert\Psi>=\Sigma_0\vert\Psi>\,{,}
 \end{equation}
 so that dynamics results reduced. A straightforward integration
 of Eqs. (\ref{64}) and (\ref{65}) gives
  \begin{equation}\label{66}
  \vert\Psi>=\vert\Omega(w)>\vert\chi(z)>\propto
  e^{i\Sigma_0z}\,e^{-i\tilde{\lambda}w^{2+s/3}}\,{,}
  \end{equation}
which is an oscillating wave function.
In the semi--classical limit, we have two first integrals
of motion: $\Sigma_0$ (i.e. the equation for $\pi_z$) and
$E_{{\cal L}}=0$, i.e. the Hamiltonian (\ref{61}) which becomes
the equation for $\pi_{w}$. Classical trajectories
in
the configuration space $\tilde{\cal Q}\equiv\{w, z\}$ are immediately
recovered
 \begin{eqnarray}
 w(t)&=&[k_1t+k_2]^{3/(s+3)}\,{,}\label{67} \\
 z(t)&=&[k_1t+k_2]^{(s+6)/(s+3)}+z_0\,{,}\label{68}
 \end{eqnarray}
then, going back to ${\cal Q}\equiv\{a, \varphi\}$,
we get the {\it classical}
cosmological behaviour
 \begin{eqnarray}
 a(t)&=&a_0(t-t_0)^{l(s)}\,{,} \label{69} \\
 \varphi(t)&=&\varphi_0(t-t_0)^{q(s)}\,{,} \label{70}
 \end{eqnarray}
 where
 \begin{equation}\label{71}
 l(s)=\frac{2s^2+9s+6}{s(s+3)}\,{,}\quad q(s)=-\frac{2s+3}{s}\,{.}
 \end{equation}
Depending on the value of $s$, we get Friedman, power--law, or
pole--like behaviours.

If we take into account generic Bianchi models, the configuration
space is ${\cal Q}\equiv \{ a_1, a_2, a_3, \varphi\}$ and more than one
symmetry can exist as it is shown in \cite{bianchi}. The
considerations on the oscillatory regime of the wave function of
the universe and the recovering of classical behaviours are
exactly the same.

\section{Fourth-Order Gravity Cosmologies}

Similar arguments work for higher--order gravity cosmology. In
particular, let us consider fourth--order gravity given by the
action
 \begin{equation}\label{72}
 {\cal A}=\int d^4x \sqrt{-g}\, f(R)\,{,}
 \end{equation}
where $f(R)$ is a generic function of scalar curvature. If
$f(R)=R+2\Lambda$, the standard second--order gravity is recovered.
We are discarding matter contributions. Reducing the action to a
point-like, FRW one, we have to write
 \begin{equation}\label{73}
 {\cal A}=\int dt {\cal L}(a, \dot{a}; R, \dot{R})\,{,}
 \end{equation}
where dot means derivative with respect to the cosmic time. The
scale factor $a$ and the Ricci scalar $R$ are the canonical
variables. This position could seem arbitrary since $R$ depends on
$a, \dot{a}, \ddot{a}$, but it is generally used in canonical
quantization \cite{vilenkin1,schmidt,lambda}. The
definition of $R$ in terms of $a, \dot{a}, \ddot{a}$ introduces a
constraint which eliminates second and higher order derivatives in
action (\ref{73}), and yields to a system of second order
differential equations in $\{a, R\}$. Action (\ref{73}) can be
written as
 \begin{equation}\label{74}
 {\cal A}=2\pi^2\int dt \left\{ a^3f(R)-\lambda\left [ R+6\left (
 \frac{\ddot{a}}{a}+\frac{\dot{a}^2}{a^2}+\frac{k}{a^2}\right)\right]\right\}\,{,}
 \end{equation}
where the Lagrange multiplier $\lambda$ is derived by varying
with respect to $R$. It is
 \begin{equation}\label{75}
 \lambda=a^3f'(R)\,{.}
 \end{equation}
 Here prime means derivative with respect to $R$. To recover a
 more strict analogy with previous scalar--tensor models, let us
 introduce the auxiliary field
 \begin{equation}\label{76}
  p\equiv f'(R)\,{,}
 \end{equation}
 so that the Lagrangian in (\ref{74}) becomes
 \begin{equation}\label{77}
 {\cal L}=6a\dot{a}^2p+6a^2\dot{a}\dot{p}-6kap-a^3W(p)\,{,}
 \end{equation}
 which is of the same form of (\ref{47}) a part the kinetic term.
This is an Helmhotz--like Lagrangian \cite{magnano} and $a, p$ are
independent fields. The potential $W(p)$ is defined as
 \begin{equation}\label{78}
 W(p)=h(p)p-r(p)\,{,}
 \end{equation}
where
 \begin{equation}\label{rh(p)}
 r(p)=\int f'(R)dR=\int pdR=f(R)\,{,} \quad h(p)=R\,{,}
 \end{equation}
such that $h=(f')^{-1}$ is the inverse function of $f'$. The
configuration space is now ${\cal Q}\equiv\{a, p\}$ and $p$ has
the same role
of the above $\varphi$. Condition (\ref{19}) is now realized by
the vector field
 \begin{equation}\label{80}
 X=\alpha (a, p)\frac{\partial}{\partial a}+\beta(a, p)
 \frac{\partial}{\partial p}+\dot{\alpha}\frac{\partial}{\partial\dot{a}}
 +\dot{\beta}\frac{\partial}{\partial\dot{p}}
 \end{equation}
 and explicitly it gives the system
 \begin{eqnarray}
&&p\left[\alpha+2a\displaystyle\frac{\partial\alpha}{\partial a}\right]p+
a\left[\beta+a\displaystyle\frac{\partial\beta}{\partial a}\right] =
 0\,{,}\label{81}  \\
&& a^2\displaystyle \frac{\partial\alpha}{\partial p}=0\,{,} \label{82} \\
&& 2\alpha+a\displaystyle \frac{\partial\alpha}{\partial
 a}+2p\displaystyle \frac{\partial\alpha}{\partial p}+
a \frac{\partial\beta}{\partial p}=0\,{,}\label{83} \\
&& 6k[\alpha p+\beta a]+a^2[3\alpha W+
a \beta \displaystyle\frac{\partial W}{\partial
 p}]=0\,{.}\label{84}
\end{eqnarray}
 The solution of this system, i.e. the existence of a Noether
symmetry, gives $\alpha$, $\beta$ and $W(p)$. It is satisfied for
\begin{equation}\label{85}
  \alpha=\alpha (a)\,{,} \qquad \beta (a, p)=\beta_0 a^sp\,{,}
 \end{equation}
where $s$ is a parameter and $\beta_0$ is an integration constant.
In particular,
 \begin{equation}\label{86}
  s=0\longrightarrow \alpha (a)=-\frac{\beta_0}{3}\, a\,{,} \quad
 \beta (p)=\beta_0\, p\,{,} \quad W(p)=W_0\, p\,{,} \quad
 k=0\,{,}
 \end{equation}
 \begin{equation}\label{87}
 s=-2\longrightarrow \alpha (a)=-\frac{\beta_0}{a}\,{,}\quad
 \beta (a, p) = \beta_0\, \frac{p}{a^2}\,{,} \quad
 W(p)=W_1p^3\,{,} \quad \forall \,\,\, k\,{,}
 \end{equation}
where $W_0$ and $W_1$ are constants. As above, the new set of
variables $Q^j=Q^j(q^i)$ adapted to the foliation induced by $X$
are given by the system (\ref{27}). Let us discuss separately the
solutions (\ref{86}) and (\ref{87}).

\subsection{The case $s=0$}

The induced change of variables
 $\displaystyle{{\cal Q}\equiv \{a, p\}
 \longrightarrow \tilde{Q}\equiv \{w, z\}}$
 can be
 \begin{equation}\label{88}
 w(a, p)=a^3p\,{,} \quad z(p)=\ln p\,{.}
 \end{equation}
 Lagrangian (\ref{77}) becomes
 \begin{equation}\label{89}
 \tilde{{\cal L}}(w, \dot{w},
 \dot{z})=\dot{z}\dot{w}-2w\dot{z}^2+\frac{\dot{w}^2}{w}-3W_0w\,{.}
 \end{equation}
 and, obviously, $z$ is the cyclic variable. The conjugate momenta are
 \begin{equation}\label{90}
 \pi_z\equiv\frac{\partial \tilde{{\cal L}}}{\partial
 \dot{z}}=\dot{w}-4\dot{z}=\Sigma_0\,{,}
 \end{equation}
\begin{equation}\label{91}
 \pi_w\equiv\frac{\partial \tilde{{\cal L}}}{\partial
 \dot{w}}=\dot{z}+2\frac{\dot{w}}{w}\,{.}
 \end{equation}
 and the Hamiltonian is
 \begin{equation}\label{92}
 {\cal H}(w, \pi_w, \pi_z)=
 \pi_w\pi_z-\frac{\pi_z^2}{w}+2w\pi^2_w+6W_0w\,{.}
 \end{equation}
 By canonical quantization, reduced dynamics is given by
 \begin{equation}\label{93}
 \left[\partial^2_z-2w^2\partial^2_w
 -w\partial_w\partial_z+6W_0w^2\right]\vert\Psi>=0\,{,}
 \end{equation}
 \begin{equation}\label{94}
 -i\partial_{z}\vert\Psi>=\Sigma_0\,\vert\Psi>\,{.}
 \end{equation}
 However, we have done simple factor ordering considerations in the
WDW equation (\ref{93}). Immediately, the wave function has an
oscillatory factor, being
 \begin{equation}\label{95}
 \vert\Psi>\sim e^{i\Sigma_0z}\vert\chi(w)>\,{.}
 \end{equation}
 The function $\vert\chi>$ satisfies the Bessel differential equation
 \begin{equation}\label{96}
 \left[w^2\partial^2_w+i\frac{\Sigma_0}{2}\,w\,\partial_w+
 \left(\frac{\Sigma_0^2}{2}-3W_0w^2\right)\right]\chi (w)=0\,{,}
 \end{equation}
 whose solutions are linear combinations of Bessel functions $Z_{\nu}(w)$
 \begin{equation}\label{97}
 \chi (w)=w^{1/2-i\Sigma_0/4}Z_{\nu}(\lambda w)\,{,}
 \end{equation}
 where
 \begin{equation}\label{98}
 \nu =\pm\frac{1}{4}\, \sqrt{4-9\Sigma_0^2-i4\Sigma_0}\,{,}
 \quad \lambda =\pm 9\sqrt{\frac{W_0}{2}}\,{.}
 \end{equation}
 The oscillatory regime for this component depends on the reality
 of $\nu$ and $\lambda$. The wave function of the universe, from Noether
symmetry (\ref{86}) is then
 \begin{equation}\label{99}
 \Psi (z, w)\sim e^{i\Sigma_0[z-(1/4)\ln w]}\,
 w^{1/2}Z_{\nu}(\lambda w)\,{.}
 \end{equation}
 For large $w$, the Bessel functions have an exponential behaviour
\cite{abramowitz}, so that the wave function (\ref{99}) can be
written as
 \begin{equation}\label{100}
 \Psi\sim e^{i[\Sigma_0z - (\Sigma_0/4)\ln w\pm \lambda w]}\,{.}
 \end{equation}
 By identifying the exponential factor of
(\ref{100}) with $S_0$, we can recover the conserved momenta
$\pi_z, \pi_w$ and select classical trajectories. Going back to
the old variables, we get the cosmological solutions
 \begin{equation}\label{101}
 a(t)=a_0e^{(\lambda/6)t}\,\exp{\left\{-\frac{z_1}{3}\,
 e^{-(2\lambda/3)t}\right\}}\,{,}
 \end{equation}
 \begin{equation}\label{102}
p(t)=p_0e^{(\lambda/6)t}\,\exp{\{z_1\,
 e^{-(2\lambda/3)t}\} }\,{,}
 \end{equation}
where $a_0, p_0$ and $z_1$ are integration constants. It is clear
that $\lambda$ plays the role of a cosmological constant and
inflationary behaviour is asymptotically recovered.

\subsection{The case $s=-2$}

The new variables adapted to the foliation for the solution
(\ref{87}) are now
 \begin{equation}\label{103}
 w(a, p)=ap\,{,}\qquad z(a)=a^2\,{.}
 \end{equation}
 and Lagrangian (\ref{77}) assumes the form
 \begin{equation}\label{104}
 \tilde{{\cal L}}(w, \dot{w}, \dot{z})=3\dot{z}\dot{w}-6kw-W_1w^3\,{,}
 \end{equation}
 The conjugate momenta are
 \begin{equation}\label{105}
 \pi_z=\frac{\partial \tilde{{\cal L}}}{\partial\dot{z}}=3\dot{w}=\Sigma_1\,{,}
 \end{equation}
 \begin{equation}\label{106}
 \pi_w=\frac{\partial \tilde{{\cal L}}}{\partial\dot{w}}=3\dot{z}\,{.}
 \end{equation}
The Hamiltonian is given by
 \begin{equation}\label{107}
 {\cal H}(w, \pi_w, \pi_z)=\frac{1}{3}\,
 \pi_z\pi_w+6kw+W_1w^3\,{.}
 \end{equation}
 Going over the same steps as above, the wave function of the
universe is given by
 \begin{equation}\label{108}
 \Psi (z, w)\sim e^{i[\Sigma_1z+9kw^2+(3W_1/4)w^4]}\,{,}
 \end{equation}
 and the classical cosmological solutions are
 \begin{equation}\label{109}
 a(t)=\pm\sqrt{h(t)}\,{,}\qquad p(t)=\pm
 \frac{c_1+(\Sigma_1/3)\,t}{\sqrt{h(t)}}\,{,}
 \end{equation}
 where
 \begin{equation}\label{h(t)}
 h(t)=\left(\frac{W_1\Sigma_1^3}{36}\right) t^4+
 \left(\frac{W_1w_1\Sigma_1}{6}\right)
 t^3+\left(k\Sigma_1+\frac{W_1w_1^2\Sigma_1}{2}\right)\,
 t^2+w_1(6k+W_1w_1^2)\, t+z_2\,{.}
 \end{equation}
 $w_1$, $z_1$ and $z_2$  are integration constants. Immediately we
 see that, for large $t$
 \begin{equation}\label{111}
 a(t)\sim t^2\,{,}\qquad p(t)\sim \frac{1}{t}\,{.}
 \end{equation}
 which is a power--law inflationary behaviour.

\section{Higher than Fourth--Order Gravity Cosmologies}

Minisuperspaces which are suitable for the above analysis can be
found for higher than fourth--order theories of gravity as
 \begin{equation}\label{112}
 {\cal A}=\int d^4x\sqrt{-g}\, f(R, \Box R)\,{.}
 \end{equation}
 In this case, the configuration space is ${\cal Q}\equiv\{a, R, \Box R\}$
 considering $\Box R$ as an independent degree of freedom
 \cite{schmidt,lambda,sixth}. The FRW point--like Lagrangian is
 formally
 \begin{equation}\label{113}
 {\cal L}={\cal L}(a, \dot{a}, R, \dot{R}, \Box R, \dot{(\Box R)})
 \end{equation}
 and the constraints
 \begin{equation}\label{114}
 R=-6\left[\frac{\ddot{a}}{a}+\left(\frac{\dot{a}}{a}\right)^2+\frac{k}{a^2}\right]\,{,}
 \end{equation}
 \begin{equation}\label{115}
 \Box R=\ddot{R}+3\frac{\dot{a}}{a}\dot{R}
 \end{equation}
 holds. Using the above Lagrange multiplier approach, we get the
Helmholtz point--like Lagrangian
 \begin{equation}\label{116}
 {\cal L}=6a\dot{a}^2p+6a^2\dot{a}\dot{p}-6kap-a^3\dot{h}q-a^3W(p,
 q)\,{,}
 \end{equation}
 where
 \begin{equation}\label{117}
 p\equiv \frac{\partial f}{\partial R}\,{,}\quad
 q\equiv \frac{\partial f}{\partial \Box R}\,{,}
 \end{equation}
 \begin{equation}\label{118}
 W(p, q)=h(p)p+g(q)q-f\,{,}
 \end{equation}
 and
 \begin{equation}\label{119}
 h(p)=R\,{,}\quad g(q)=\Box R\,{,}\quad f=f(R,\Box R){.}
 \end{equation}
 Now the minisuperspace is three--dimensional but, again, the
Noether symmetries can be recovered. Cases of physical interest
\cite{sixth} are
 \begin{equation}\label{120}
 f(R, \Box R)=F_0R+F_1R^2+F_2R\Box R\,{,}
 \end{equation}
 \begin{equation}\label{121}
 f(R, \Box R)=F_0R+F_1\sqrt{R\Box R}\,{,}
 \end{equation}
 discussed in details in \cite{oneloop}.  Also here the existence
of the symmetry selects the form of the model and allows to reduce the
dynamics. Once it is identified, we can perform the change of
variables induced by foliation using Eqs. (\ref{27}), if a
symmetry is present, or Eqs. (\ref{29}), is two symmetries are
present. In both cases,
 \begin{equation}\label{122}
 {\cal Q}\equiv \{a, R, \Box R\}\longrightarrow \tilde{\cal Q}\equiv
 \{z, u, w\}\,{,}
 \end{equation}
 where one or two variables are cyclic in Lagrangian (\ref{116}).
Taking into account, for example, the case (\ref{121}), we get
 \begin{equation}\label{123}
 \tilde{{\cal
 L}}=3[w\dot{w}^2-kw]-F_1\left[3w\dot{w}^2u+3w^2\dot{w}\dot{u}+
 \frac{w^3\dot{z}\dot{u}}{2u^2}-3kwu\right]\,{,}
 \end{equation}
 where we assume $F_0=-1/2$, the standard Einstein coupling, $z$ is the
cyclic variable and
 \begin{equation}\label{124}
 z=R\,{,}\quad u=\sqrt{\frac{\Box R}{R}}\,{,}\quad w=a\,{.}
 \end{equation}
 The conserved quantity is
 \begin{equation}\label{125}
 \Sigma_0=\frac{w^3\dot{u}}{2u^2}\,{.}
 \end{equation}
 Using the canonical procedure of quantization and deriving the
WDW equation from (\ref{123}), the wave function of the universe
is
 \begin{equation}\label{126}
\vert \Psi>\sim e^{i\Sigma_0z}\vert\chi(u)>\vert\Theta(w)>\,{,}
 \end{equation}
 where $\chi(u)$ and $\Theta(w)$ are combinations of Bessel
functions. The oscillatory subset of the solution is evident.
In the semi--classical limit,
using the conserved momentum (\ref{125}), we obtain the
cosmological behaviours
 \begin{equation}\label{127}
 a(t)=a_0t\,{,} \quad a(t)=a_0t^{1/2}\,{,}\quad
 a(t)=a_0e^{k_0t}\,{,}
 \end{equation}
 depending on the choice of boundary conditions.

\section{Discussion and Conclusions}

In this paper, we discussed the connection of Noether symmetries
for minisuperspace cosmological models to the recovering of
classical solutions. If the wave function of the universe is
related to the probability to get a given classical cosmology, the
existence of such symmetries tell us when the WDW wave function of the
universe has oscillatory behaviours connected to the recovering
of correlations between coordinates and conjugate canonical momenta
\cite{halliwell}. I this sense, the Hartle criterion to get correlations
capable of selecting classical universes works.

Some remarks are necessary at this point. First of all, we have to
stress that the wave function is {\it only} related to the
probability to get a certain behaviour but it is not the
probability amplitude since, till now, quantum cosmology is not a
unitary theory. Furthermore, the Hartle criterion works in the
context of an Everett--type interpretation of quantum cosmology
\cite{everett,finkelstein} which assumes the ideas that the
universe branches into a large number of copies of itself whenever
a measurement is made. This point of view is called {\it Many
Worlds} interpretation of quantum cosmology.
Such an interpretation is just one way
of thinking and gives a formulation of quantum mechanics designed
to deal with correlations internal to individual, isolated systems.
The Hartle criterion gives an operative interpretation of such
correlations. In particular, if the wave function is {\it strongly
peaked} in some region of configuration space, we predict that we
will observe the correlations which characterize that region. On
the other hand, if the wave function is {\it smooth} in some
region, we predict that correlations which characterize that
region are precluded to the observations.

If the wave function is neither peaked nor smooth, no predictions
are possible from observations. In other words, we can read the
{\it correlations} of some region of minisuperspace as {\it causal
connections}. However, the validity of minisuperspace approximation is
often not completely accepted and it is still matter of debate
\cite{kuchar}.

As we said above, the analogy with standard quantum mechanics is
straightforward. By considering the case in which the individual
system consists of a large number of identical subsystems, one can
derive from the above interpretation, the usual probabilistic
interpretation of quantum mechanics for the subsystems
\cite{hartle1,halliwell}.

What we proposed in this paper is a criterion by which the Hartle
point of view can be recovered without arbitrariness. If a Noether
symmetry (or more than one) is present for a given minisuperspace
model, then oscillatory subsets of the wave
function of the universe are found. Viceversa, oscillatory parts
of the wave function can be always connected to conserved momenta
and then to Noether symmetries.

>From a general point of view, this is the same philosophy of many
branches of physics: Finding symmetries allows to solve dynamics,
gives the main features of systems and simplify the interpretation
of results.

However the above scheme should be enlarged to
more general classes of minisuperspaces in order to seek for its
application to the full field theory (i.e. to the infinite--dimensional
superspace). Only in this sense, one could claim for the validity of
the approach to the full semiclassical limit of quantum cosmology.

\end{document}